# The soft x-ray imager on THESEUS: the transient high-energy survey and early universe surveyor

Paul O'Brien[a,1], Ian Hutchinson[a], Hannah Lerman[a], Charlotte Feldman[a], Melissa McHugh[a], Alex Lodge[a], Richard Willingale[a], Andy Beardmore[a], Roisin Speight[a], Paul Drumm[a]

[a]School of Physics and Astronomy, University of Leicester, University Road, Leicester, LE1 7RH, UK

## ABSTRACT

We are entering a new era for high energy astrophysics with the use of new technology to increase our ability to both survey and monitor the sky. The Soft X-ray Imager (SXI) instrument on the THESEUS mission will revolutionize transient astronomy by using wide-field focusing optics to increase the sensitivity to fast transients by several orders of magnitude. The THESEUS mission is under Phase A study by ESA for its M5 opportunity. THESEUS will carry two large area monitors utilizing Lobster-eye (the SXI instrument) and coded-mask (the XGIS instrument) technologies, and an optical-IR telescope to provide source redshifts using multi-band imaging and spectroscopy. The SXI will operate in the soft (0.3-5 keV) X-ray band, and consists of two identical modules, each comprising 64 Micro Pore Optics and 8 large-format CMOS detectors. It will image a total field of view of 0.5 steradian instantaneously while providing arcminute localization accuracy. During the mission, the SXI will find many hundreds of transients per year, facilitating an exploration of the earliest phase of star formation and comes at a time when multi-messenger astronomy has begun to provide a new window on the universe. THESEUS will also provide key targets for other observing facilities, such as Athena and 30m class ground-based telescopes.

## 1. INTRODUCTION

THESEUS – the Transient High Energy Survey and Early Universe Surveyor – is a candidate M5 mission under Phase A study by the European Space Agency (ESA)[1,2]. The primary scientific goals of THESEUS are to address the Early Universe ESA Cosmic Vision theme "How did the Universe originate and what is made of?" and in particular sub-themes 4.1 "Early Universe", 4.2 "The Universe taking shape" and 4.3 "The evolving violent Universe". THESEUS will also impact on the "The Gravitational Wave Universe" and "The Hot and Energetic Universe" themes. Using a combination of two wide-field monitor instruments[3] and an optical/IR telescope[4] mounted on a fast-slewing spacecraft, THESEUS will find several thousand transients during its prime mission. Alerts will be rapidly communicated to ground and uploads will also be used to follow-up alerts from other facilities. THESEUS will have unique scientific capabilities to: a) Explore the Early Universe (the cosmic dawn and re-ionization era) by unveiling the Gamma-Ray Burst (GRB) population in the first billion years; and b) Perform an unprecedented deep monitoring of the soft X-ray transient Universe, thus providing a crucial synergy with next-generation gravitational waves and neutrino detectors (multi-messenger astrophysics), as well as the large next-generation electromagnetic facilities. Here we describe the Soft X-ray Imager (SXI) instrument, one of the two wide-field monitors on THESEUS. The SXI is being developed by a consortium of scientists from ESA member states, led by the UK, including Belgium, Spain, Switzerland and the Czech Republic.

## 2. SOFT X-RAY IMAGER

**Overview**

The SXI instrument comprises 2 identical modules (see Figure 1 and Table 1). Each module is a wide-field, "Lobster eye" X-ray telescope, using the X-ray imaging principle first described by Angel[3]. This configuration provides for a wide FOV,

---

focusing X-ray imaging system with an effective area maintained across the entire FOV. The optics aperture for the SXI modules is formed by an array of 8x8 square micro pore optics (MPOs) mounted on a spherical frame with a radius of curvature of 600 mm[6]. Figure 2 shows the point spread function. X-rays which reflect off the square pore sides form a central focus (even number of reflections) or a line focus (odd number), giving a cross-arm PSF. In this configuration some 75% of the incident X-rays are focused, with detailed simulations showing the optics provide the required <2 arc-minutes transient location accuracy. The off-axis angle at which the cross arms first go to zero is determined by the L/d ratio of the pores, where L is the MPO thickness and d the pore width. For optimum performance (combination of sensitivity and imaging quality) at 1 keV we require L/d=60, and use L=2.4mm and d=40 µm. Mounted directly onto the frame, behind the optics, are a set of magnets forming an electron diverter. Optics heaters are also mounted on the frame, and together these components comprise the optics assembly, which provides a FOV of ~0.25 steradians (31x31 degrees) per module. The 2 modules will be aligned on the sky parallel to the spacecraft Sun shade, with a 1-degree wide overlap, coaligned with the IRT FOV to provide redundancy. The SXI modules comprise an optic assembly focusing light onto a focal plane assembly comprising 8 CMOS detectors[7]. A dedicated instrument structure comprising of an Aluminium Optic Assembly Frame, a tapered Aluminium Telescope tube and an Aluminium/Titanium Instrument support structure to maintain the focal length between the optics and the detectors and provide the interface with the spacecraft platform.

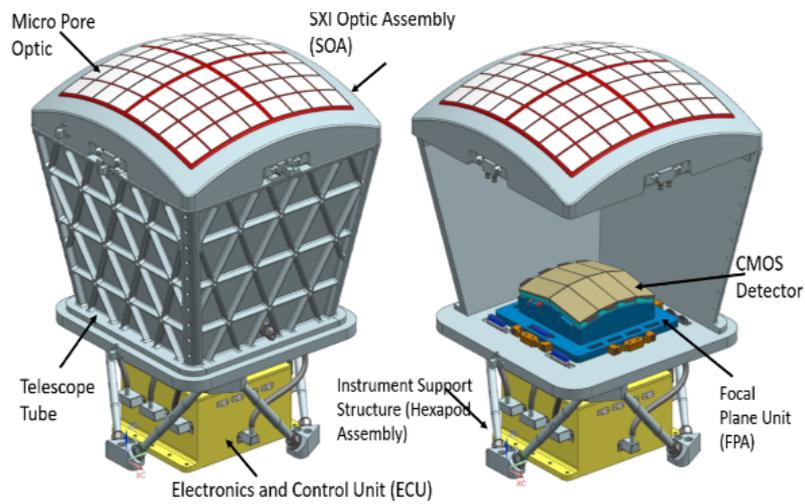

Figure 1. left panel shows the SXI module exterior view; right panel cut-away shows the focal plane assembly located below the optics assembly. The electronics box is below the focal plane within the structure which mounts to the spacecraft.

Table 1. SXI module physical characteristics.

| Energy band (keV) | 0.3-5 |
|---|---|
| Optics configuration | 8x8 square pore MPOs |
| MPO size (mm$^2$) | 40x40 |
| Focal length (mm) | 300 |
| Focal plane detectors | CMOS array |
| CMOS size (mm$^2$) | 80x40 |
| CMOS pixel size (µm) | 40 |
| CMOS pixel Number | 2000x1000 |
| Number of CMOS | 8 |
| Module Field of View (sr) | 0.25 |
| Centroiding accuracy (best, worst) (arcsec) | (<30, 180) |

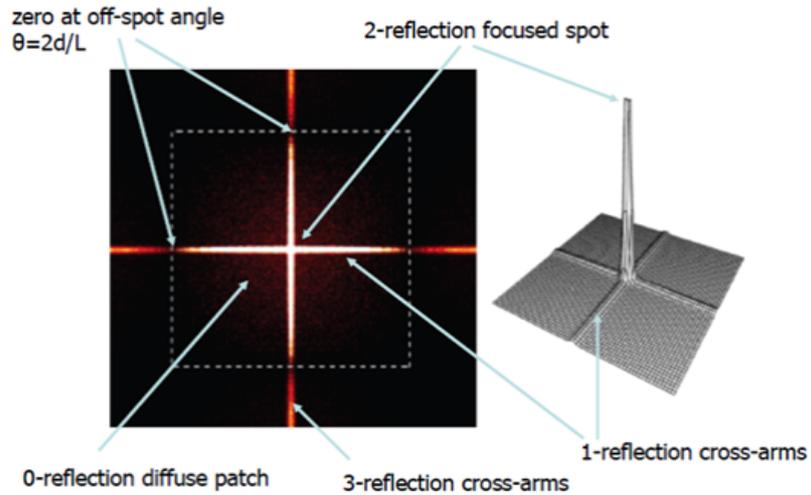

Figure 2 The MPO point spread function showing the cross-arm shape. The inner dotted square shows where the PSF cross-arms first go to zero.

The SXI module is enclosed in MLI in order to maintain thermal control. A thermal control schematic is shown in Figure 3. The MPOs and detectors both have sky-facing thin Aluminium layers acting as a light block. The MPO Aluminium layer also efficiently rejects Earth light to aid thermal control. The SXI module has two key thermal control challenges: maintaining the MPOs within their optimum temperature range (20°C to 30°C) with a small gradient, and maintaining the CMOS detectors at their optimum temperature (<=-30°C) and stability (±1.5°C). The former is achieved through PID controlled heaters installed on the Optic Frame; the latter is achieved via a spacecraft provided Cold Finger (CF), the use of TECs, and good thermal coupling between the FPA support (cold base) and the warm base which is part of the instrument support structure (see Figure 4).

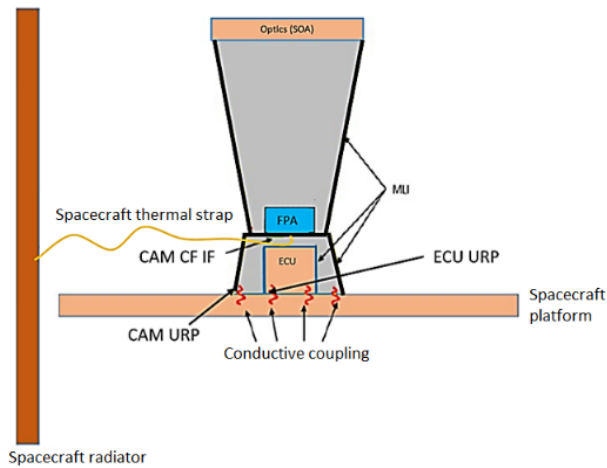

Figure 3. A Schematic of the SXI thermal concept showing the module URP couplings (the ECU and instrument support structure) and CF coupling (the cold base).

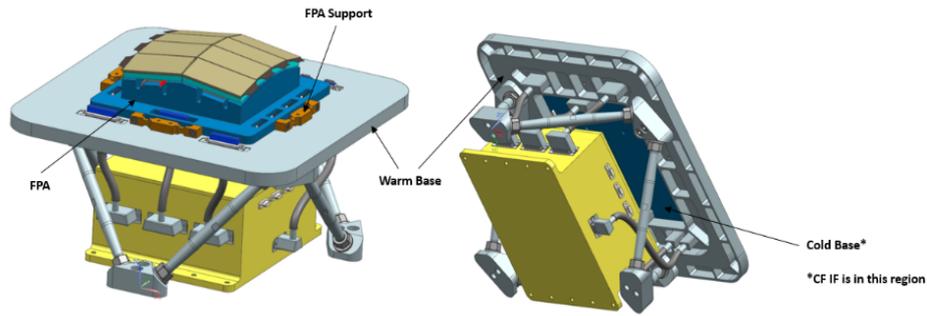

Figure 4. Focal plane assembly cold base and warm base.

The Phase A mechanical analysis, performed with a simple finite element (FE) model, demonstrates a first mode (Figure 5) within the anticipated launch specification and material limits.

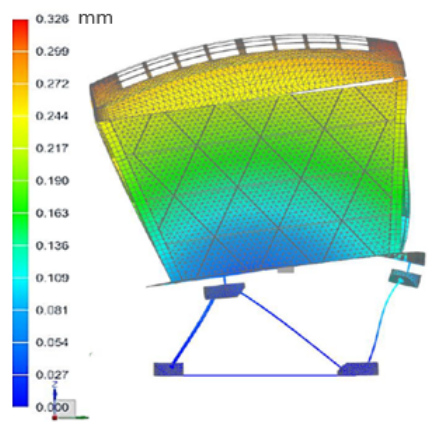

Figure 5. Example of the mechanical analysis (first mode) performed on an SXI module.

A functional block diagram of the SXI is shown in Figure 6. The SXI modules utilizes front-end electronics (FEE) for focal plane control/data acquisition, and back-end electronics (BEE) for thermal control and housekeeping. The FEE provide local power conditioning for the detector bias lines and clocks, and routes the digital data output from the detectors to a small processor for event list generation – data is transferred between the FEE and the Data Handling Unit (DHU) via SpaceWire interface.

There are 4 detector pairs (per instrument module), and 1 FEE board will be required to drive 2 of these detector pairs. The 2 FEE PCBs and a power board are housed within an Electronics and Control Unit (ECU) for each module that is placed in close proximity to the focal plane (the ECU is shown underneath the focal plane assembly in Figure 1 and Figure 4). The BEE are located inside the SXI DHU box (which can be placed up to 3m away from the modules) and provide overall thermal control for both SXI modules (i.e., TEC drive/control for the focal plane and heater drive/control for the optics structure) as well as handling of the housekeeping data (thermal sensors on the optics structure, focal plane, and ECU box). The SXI DHU provides the data interface to the spacecraft platform and controls (switches) the spacecraft power to the SXI ECU.

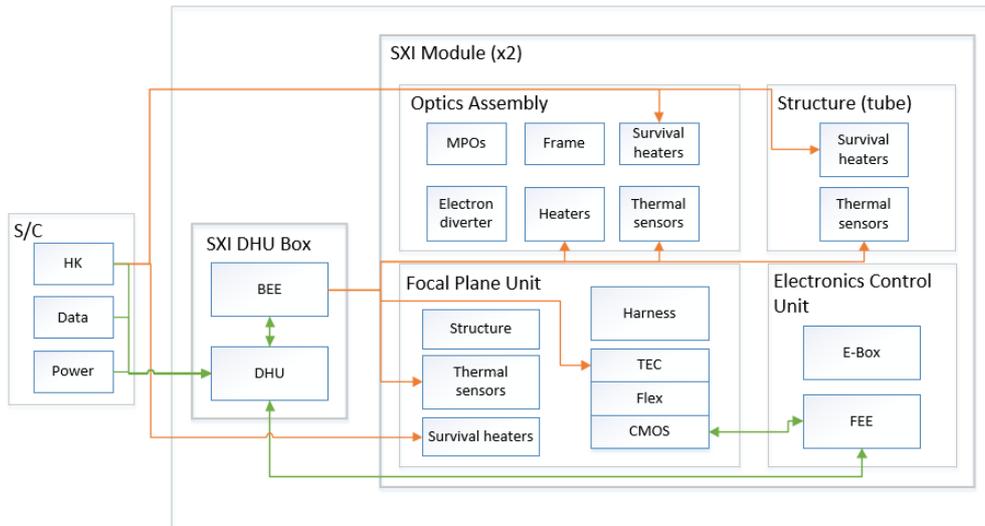

Figure 6. SXI block diagram.

## 2.1 Interfaces and resource requirements

Each SXI module has an estimated mass of 35.9kg. The mechanical and thermal interfaces with the spacecraft are via CAM CF (for SC thermal strap) and via ECU and Instrument Support Structure mechanical interfaces with the PLM, see Figure 7. Each module will also require a purge interface with the spacecraft. The total power for each SXI module is ~53W. The SXI DHU requires ~10W of power and hence the total power requirement for the full SXI instrument (i.e. 2x modules and the DHU) is <120W.

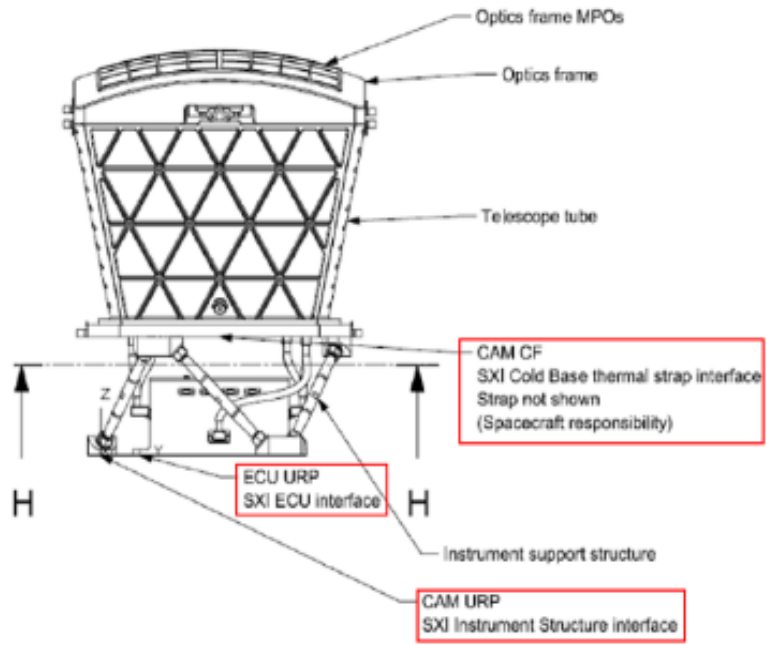

Figure 7. SXI module thermal interfaces.

**2.2 Operational requirements**

The SXI instrument operates almost always in the same observing mode, except when performing calibration, which is performed in parallel across the module detectors. The detectors are clocked out every 100ms, photons are graded in the electronics and the photon stream (with grade, energy and pixel locations) are passed to the DHU within which runs the on-board trigger system which searches for transients. All photons are telemetered to ground to conduct the X-ray sky survey. Event list processing (e.g., trigger algorithms) occur in the DHU which also provides event list and housekeeping data storage. The SXI telemetry rates for science mode average 100kbit/s; while for house-keeping they average 12.8kbit/s.

**2.3 Heritage and technology development**

The THESEUS SXI optics are based on those developed for the recently launched ESA BepiColombo mission, and also those under construction for the China-France SVOM and ESA-China SMILE missions. The SXI optic is a larger version of the SVOM MXT optic, the flight model for which has been assembled and has undergone thermal and vibration testing (Figure 8).

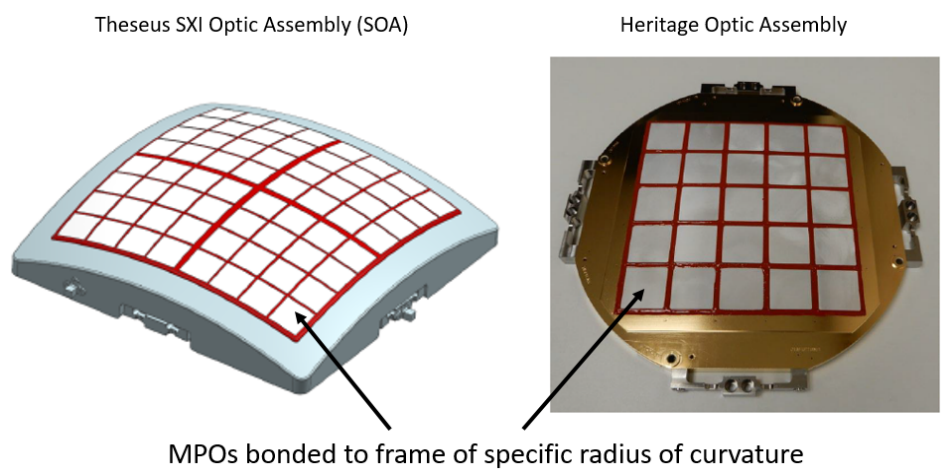

Figure 8. THESEUS SXI optics assembly compared to the SVOM MXT optics assembly (CNES).

The SVOM and SMILE optics assemblies use the same materials as baselined for the THESEUS SXI, and these missions share a common MPO bonding, testing and alignment process. These missions all make use of MPOs supplied by the Photonis France SAS company, but during Phase A the possible use of an alternative MPO supplier – the Chinese North Night Vision Technology (NNVT) company, which is presently supplying MPOs for the Chinese Einstein Probe mission – has been investigated. An ESA funded TDA is underway to compare the MPOs from both potential suppliers, and enables MPO development at Photonis to further improve optics quality, hence reducing risk and cost for MPO provision.

There are two ongoing detector technology development studies involving large-format CMOS X-ray that are relevant to the SXI: an ESA funded TDE activity and an ESA funded TDA activity. The TDE activity involves the development of a large area CMOS detector that is suitable for X-ray spectroscopy applications, i.e., a CMOS device with large, deep depleted pixels (~35-40μm square, to optimally match X-ray event size) and appropriate on-chip optical blocking filters (e.g. Al, to ensure suitable rejection of straylight whilst maintaining soft X-ray detection efficiency). A detailed study of the X-ray performance of the CMOS devices developed during the TDE activity will be performed during the SXI detector TDA. The scope of the TDA also includes the top-level design of the SXI camera system (including thermal and mechanical analysis, top level design of the FEE and BEE, and development of an autonomous event detection algorithm) and a full, end-to-end demonstration of the SXI system level performance (including representative FEE and BEE systems, and representative optics).

## 2.4 Performance

The SXI optics PSF is characterized by a central spot and four extended cross-arms. An optimized event detection algorithm which uses different fractions of the detailed shape of the PSF as a function of trigger time will be used to ensure that as many of the X-ray sources as possible (i.e., that are located within the FoV of the instrument) activate the trigger system. The X-ray events associated with the detected source will then be stored in the instrument's DHU. Ideally, the onboard algorithm would perform matched filtering on the full PSF, but in order to reduce computation time and in order to minimize the effects of the sky and detector backgrounds (i.e., maximize the source to background flux ratio of the instrument), the algorithm utilizes a cross-beam ROI shape that is limited to a fraction of the overall source counts (i.e., the event extraction region is limited to specific regions around the central core and arms).

A schematic of a typical Lobster eye cross-beam (Figure 2) representing the PSF is shown in Figure 9. The parameters B and a depend on the quality of the optics, and can be optimized with exposure time such that the cross-beam contains as high a fraction of source counts as possible while limiting the contributions from the background. H is the distance to the first minimum in the arms. If the parameters are set such that the source flux is limited to 50% of the total, B is a robust measure of the half energy width of the PSF.

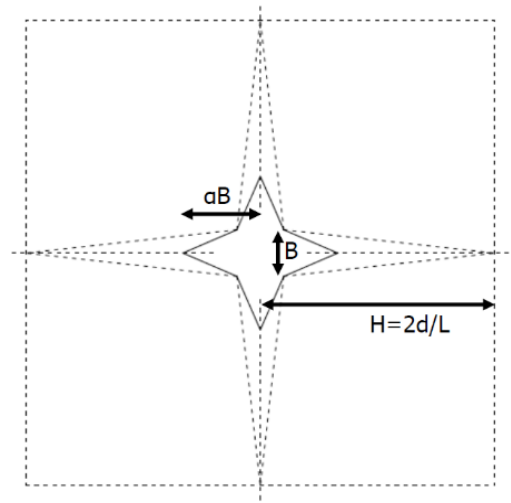

Figure 9. A schematic of the Lobster-eye cross-beam shape.

On board an accurate and rapid source position is determined by implementing a multi-stage centroiding process. Initially, a centre of gravity calculation is performed for all events found within rectangular zones on the detector image (created by stacking many individual frames together). In the second stage, the centroid position is further refined by performing sequential, single axis binning and weighted centroiding (based on the known profile of the PSF). In the final stage of the process, the parameterized PSF ROI described above is used to obtain a final determination for the source position. This is then compared to an on-board list of known sources to determine if a new potential target of interest has been found which may result in a trigger possibly requiring a spacecraft slew and a communication to the ground. On board comparison between sources found by both monitors can also be performed.

Ray tracing simulations across the 0.3-5keV energy range show that the SXI effective area peaks at 1keV (see Figure 10) and that the optics exhibit a FWHM of ~6 arcmins at the peak energy. The curve in Figure 10 shows the SXI half total effective area (accounting for the Aluminium filter on the optics, an Aluminium coating on the detector and the quantum efficiency of the CMOS device) achieved when using the 50% cross-beam extraction algorithm. In this case, source counts are collected across a total area of 675 square arcmins (i.e., equivalent to a circular beam of radius 14.7 arcmin).

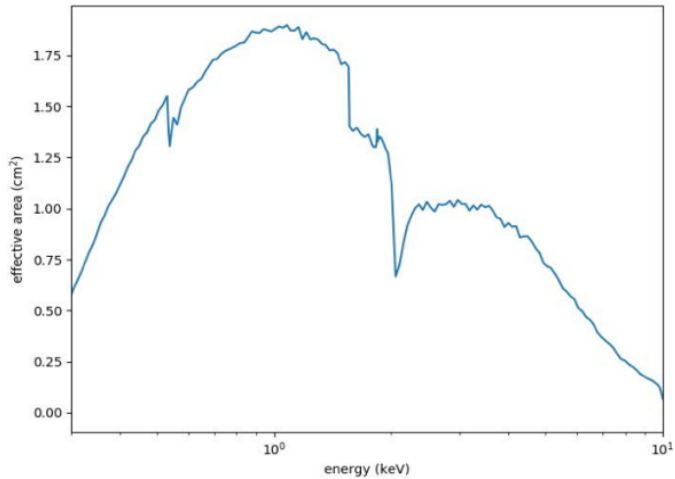

Figure 10. SXI half total effective area.

Using the effective area shown in Figure 10 combined with an estimate of the particle and sky background rates, the minimum detectable flux level and the localization accuracy achieved as a function of time can be determined. Here we assume a source with a powerlaw spectrum, photon index 1.7, absorbing column of $5 \times 10^{20}$ cm$^2$, sky background of $1.14 \times 10^{-5}$ cts s$^{-1}$ arcmin$^2$, and particle background of $9.75 \times 10^{-7}$ cts s$^{-1}$ arcmin$^2$.

To demonstrate the localization accuracy achievable with the SXI instrument as a function of time, a Monte Carlo simulation of this approach was generated to model the interactions of the photons imaged by the SXI instrument (i.e., by modelling the expected SXI PSF shape and interactions with the detector, and applying the cross-beam analysis method for centroiding). Through this model the minimum number of counts required to achieve the localization accuracy at a subpixel level was determined. Figure 11 shows the centroiding accuracy as a function of photon counts for a weighted centroiding method. Through the analysis, it has been demonstrated that when using a standard weighted centroiding method, the position of a source can be determined from 50 counts to an accuracy of less than 1 pixel.

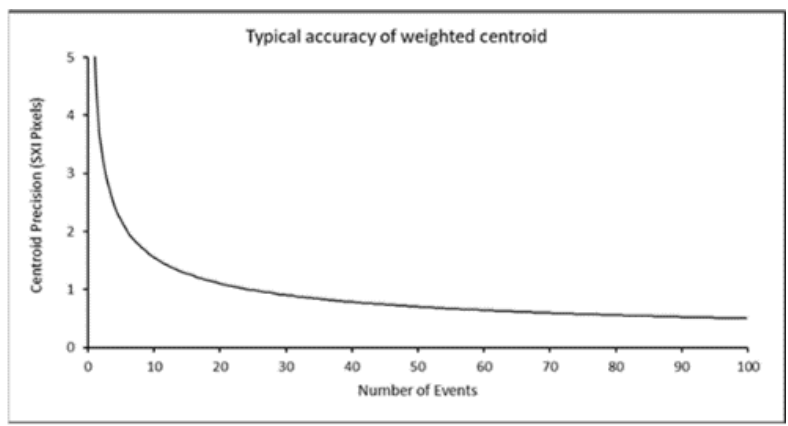

Figure 11. The typical accuracy of a weighted centroid as a function of photon events.

## ACKNOWLEDGEMENTS

We acknowledge contributions from the SXI consortium and the wider THESEUS team to the development of the SXI. We also acknowledge funding from ESA and several ESA member states, including from the UK Space Agency.

## REFERENCES

[1] Amati, L., O'Brien, P., Götz, D., and Bozzo, E., "The transient high-energy sky and early universe surveyor," Proc. of SPIE 11444-302 (2020)
[2] Amati, L., Bozzo, E., O'Brien, P., and Götz, D., "The transient high-energy sky and early universe surveyor (theseus)," Mem. S.A.It. 75, 282 (2019)
[3] Labanti, C., Campana, R., Fuschino, F., and Amati, L., "The x/gamma-ray imaging spectrometer (xgis) on-board theseus: Design, main characteristics, and concept of operation," Proc. of SPIE 11444-303 (2020)
[4] Götz, D., Basa, S., Bozzo, E., and Tenzer, C., "The infra-red telescope on board the theseus mission," Proc. of SPIE 11444-305 (2020)
[5] Angel, J. R. P., "Lobster eyes as X-ray Telescopes", Ap. J. **233**, 364-373 (1979)
[6] Feldman, C., O'Brien, P., Willingale, D., et al., "The development of the THESEUS SXI optics", Proc. of SPIE 11444-95 (2020)
[7] McHugh, M., et al., ""Development of an imaging system for the THESEUS SXI instrument", Proc. Of SPIE 11444-284 (2020)